\documentclass[twocolumn,nofootinbib]{revtex4-2}
\usepackage{amsmath,amssymb,bm,graphicx,hyperref}
\allowdisplaybreaks[1]

\renewcommand\d{\partial}
\newcommand\eps{\varepsilon}
\newcommand\E{{\bm{E}}}
\newcommand\e{{\bm{e}}}
\renewcommand\k{{\bm{k}}}
\newcommand\x{{\bm{x}}}
\newcommand\A{\mathcal{A}}
\newcommand\B{\mathcal{B}}
\renewcommand\L{\mathcal{L}}

\begin{document}

\title{Optical Magnus effect on gravitational lensing}

\author{Yusuke Nishida}
\affiliation{Department of Physics, Institute of Science Tokyo,
Ookayama, Meguro, Tokyo 152-8551, Japan}

\date{March 2026}

\begin{abstract}
The optical Magnus effect refers to a transverse shift of a trajectory of light caused by its polarization and arises as a correction to geometrical optics at the linear order in wavelength.
Here, we start from Maxwell's equations in a curved spacetime to derive the equations of motion for a wave packet of circularly polarized light, which confirm the known results involving the helicity-dependent anomalous velocity with some generalization and clarification.
We then study possible consequences of the optical Magnus effect on gravitational lensing in the Schwarzschild spacetime as well as in an expanding spacetime with a weak gravitational potential.
Among others, by formulating a lens equation modified to incorporate the optical Magnus effect, the Einstein ring is found impossible to emerge from a point source for any axially symmetric thin lens.
Analytic solutions to the modified lens equation are also obtained for simple lens models, illuminating how image formation is affected by the optical Magnus effect.
\end{abstract}

\maketitle

\section{Introduction}
Polarization has been an important probe in high-precision astrophysics and cosmology, providing us with new insights unavailable from intensity measurements alone.
Among others, the $B$-mode polarization of the cosmic microwave background is useful to constrain models of cosmic inflation and primordial gravitational waves because it cannot be produced by density fluctuations~\cite{Seljak:1997a,Seljak:1997b,Kamionkowski:1997,Kamionkowski:2016}.
The polarization images taken by the Event Horizon Telescope and the Imaging X-ray Polarimetry Explorer already brought out information on structures and dynamics of magnetized plasmas around supermassive black holes~\cite{Akiyama:2021a,Akiyama:2021b,Liodakis:2022}.
Furthermore, possible polarization modes of gravitational waves are expected to serve as diagnostics to test alternative theories of gravity~\cite{Eardley:1973a,Eardley:1973b,Will:2014}.
The polarization of light is typically considered to be parallel transported from a source to an observer, with its trajectory following a null geodesic of the background spacetime independent of the wavelength and polarization of light.
This is an assumption relying on the geometrical-optics approximation that is valid in the limit of negligible wavelength~\cite{Misner:1973,Schneider:1992}.

However, the trajectory of light is to acquire a dependence on its polarization when the wave nature of light is incorporated at the linear order in wavelength.
This phenomenon in classical optics is known as the optical Magnus effect or the spin Hall effect of light~\cite{Dooghin:1992,Liberman:1992,Bliokh:2004a,Bliokh:2004b,Onoda:2004,Bliokh:2015,Ling:2017}.
It manifests itself as a transverse shift of a trajectory of circularly polarized light when it propagates through an inhomogeneous optical medium.
Because a spacetime with a static gravitational potential can be modeled as an optical medium with a spatially varying refractive index~\cite{Felice:1971}, a gravitational analog of the optical Magnus effect is expected and has indeed been derived with various approaches such as the Foldy-Wouthuysen transformation~\cite{Berard:2006,Gosselin:2007}, the Wentzel-Kramers-Brillouin approximation~\cite{Frolov:2011,Yoo:2012}, the path integral formulation~\cite{Yamamoto:2017,Yamamoto:2018}, and the Mathisson-Papapetrou-Dixon equations~\cite{Duval:2017,Duval:2019}.
See also Refs.~\cite{Oancea:2020,Frolov:2020,Andersson:2021,Harte:2022} for recent developments as well as Refs.~\cite{Oancea:2019,Andersson:2023} for reviews on this subject.

The purpose of our work is to study possible consequences of the optical Magnus effect on gravitational lensing, which is the deflection of light by a massive object predicted by Einstein's theory of general relativity~\cite{Misner:1973,Schneider:1992}.
In order to make our work self-contained, we first derive the equations of motion for a wave packet of circularly polarized light starting from Maxwell's equations in a curved spacetime in Sec.~\ref{sec:optical}.
We then apply the resulting equations of motion to gravitational lensing in the Schwarzschild spacetime in Sec.~\ref{sec:schwarzschild} as well as in an expanding spacetime with a weak gravitational potential in Sec.~\ref{sec:weak}.
Finally, our findings are summarized in Sec.~\ref{sec:summary} together with some outlook.
We note that the speed of light is set to $c=1$ in Sec.~\ref{sec:optical} and the Appendixes, whereas $c$ is kept explicitly in the other sections, and the summation over repeated indices is assumed regardless of their positions.

\section{Optical Magnus effect}\label{sec:optical}
\subsection{Maxwell's equations}
In order to derive the gravitational analog of the optical Magnus effect, we start from Maxwell's equations in a curved spacetime,
\begin{align}\label{eq:maxwell}
\nabla_{\!a}F^{ab} = 0, \qquad \nabla_{\!a}\tilde{F}^{ab} = 0,
\end{align}
where $F^{ab}$ is the electromagnetic tensor and $\tilde{F}^{ab}=-\epsilon^{abcd}F_{cd}/2$ is its dual with $\epsilon^{0123}=-1/\sqrt{-g}$.
The metric in the foliated spacetime is decomposed as
\begin{align}
ds^2 = -\alpha^2dt^2 + \gamma_{ij}(dx^i + \beta^idt)(dx^j + \beta^jdt)
\end{align}
in terms of the lapse function $\alpha$, the shift vector $\beta^i$, and the spatial metric $\gamma_{ij}$ on the spacelike hypersurface $\Sigma_t$.
The timelike unit vector normal to $\Sigma_t$ reads $n^a=(1/\alpha,-\beta^i/\alpha)$, and thus, $n_a=(-\alpha,0,0,0)$.
The electric and magnetic fields measured by an Eulerian observer are then provided by
\begin{align}
E^a = F^{ab}n_b, \qquad B^a = \tilde{F}^{ab}n_b.
\end{align}
Because of $n_aE^a=n_aB^a=0$, the electromagnetic tensor and its dual can be expressed as
\begin{align}
F^{ab} &= n^aE^b - n^bE^a + \epsilon^{abc}B_c, \\
\tilde{F}^{ab} &= n^aB^b - n^bB^a - \epsilon^{abc}E_c,
\end{align}
with $\epsilon^{abc}\equiv n_d\epsilon^{dabc}$.

Maxwell's equations in Eq.~(\ref{eq:maxwell}) are projected along $n_a$ by $n_b\nabla_{\!a}F^{ab}=0$, leading to
\begin{align}\label{eq:gauss}
D_iE^i = 0, \qquad D_iB^i = 0,
\end{align}
where $D_i$ is the covariant derivative compatible with $\gamma_{ij}$.
On the other hand, they are projected onto $\Sigma_t$ by $({\delta^i}_b+n^in_b)\nabla_{\!a}F^{ab}=0$, leading to
\begin{align}
\d_tE^i &= (\L_\beta + \alpha K)E^i + \epsilon^{ijk}D_j(\alpha B_k), \label{eq:ampere}\\
\d_tB^i &= (\L_\beta + \alpha K)B^i - \epsilon^{ijk}D_j(\alpha E_k), \label{eq:faraday}
\end{align}
where $\L_\beta E^i=\beta^jD_jE^i-E^jD_j\beta^i$ is the Lie derivative along the shift vector and $K=-\nabla_{\!a}n^a$ is the trace of the extrinsic curvature tensor.
See Refs.~\cite{Thorne:1982,Alcubierre:2009,Baumgarte:2010} for further details.
Obviously, Eqs.~(\ref{eq:gauss}), (\ref{eq:ampere}), and (\ref{eq:faraday}) represent Gauss's law, Amp\`ere's law, and Faraday's law in a curved spacetime, respectively.

We now assume an irrotational spacetime with $\beta^i=0$ and introduce a densitized complex field as $\Psi^i\equiv\sqrt\gamma\,(E^i+iB^i)/\sqrt{2}$.
By employing $\alpha K=-\d_t\ln\!\sqrt\gamma+D_j\beta^j$ and the Levi-Civita symbol $\eps^{ijk}=\sqrt\gamma\,\epsilon^{ijk}\in\{0,\pm1\}$, each pair of equations for the electric and magnetic fields can be combined into
\begin{align}
\d_i\Psi^i &= 0, \\
\d_t\Psi^i &= -i\eps^{ijk}\d_j\!\left(\frac{\alpha}{\sqrt\gamma}\Psi_k\right).
\end{align}
The second equation leads to the continuity equation,
\begin{align}
\d_t\rho + \d_ij^i
= \d_t\!\left(\frac{\alpha}{\sqrt\gamma}\gamma_{ij}\right)\Psi^{i*}\Psi^j,
\end{align}
where $\Psi^{i*}$ is the complex conjugate of $\Psi^i$ and
\begin{align}
\rho = \frac{\alpha}{\sqrt\gamma}\gamma_{ij}\Psi^{i*}\Psi^j, \qquad
j^i = -i\eps^{ijk}\frac{\alpha^2}{\gamma}\Psi_j^*\Psi_k
\end{align}
are proportional to the energy density and the Poynting vector, respectively.
Therefore, the total energy of the electromagnetic field is conserved only when $\alpha\gamma_{ij}/\sqrt\gamma$ is independent of time.
For example, it is achieved for $\alpha=a(t)\sqrt{A(\x)}$ and $\gamma_{ij}=a(t)^2C(\x)\delta_{ij}$, which is the case considered in the following discussion.

\subsection{Wave packet motion}\label{sec:motion}
In the background spacetime described by
\begin{align}\label{eq:metric}
ds^2 = a(t)^2[-A(\x)dt^2 + C(\x)\delta_{ij}dx^idx^j],
\end{align}
Maxwell's equations are thus expressed in the Dirac-like form of
\begin{align}
\d_i\Psi^i(t,\x) &= 0, \label{eq:constraint}\\
i\d_t\Psi^i(t,\x) &= \eps^{ijk}\d_j[v(\x)\Psi^k(t,\x)]. \label{eq:evolution}
\end{align}
Here, $a(t)$ is the scale factor, $t$ is the conformal time, $\x$ are comoving coordinates, and $v(\x)\equiv\sqrt{A(\x)/C(\x)}$ is an effective velocity.
We then introduce the Fourier representation of the complex electromagnetic field,
\begin{align}\label{eq:fourier}
\bm\Psi(t,\x) = \int\!\frac{d\k}{(2\pi)^3}\psi(t,\k)\e(\k)e^{\lambda i\k\cdot\x},
\end{align}
where $\e(\k)\equiv[\hat\e_1(\k)+i\hat\e_2(\k)]/\sqrt2$ is a complex polarization vector defined in terms of three mutually orthogonal unit vectors $\hat\k$, $\hat\e_1(\k)$, $\hat\e_2(\k)$, satisfying $\hat\k=\hat\e_1(\k)\times\hat\e_2(\k)$.
Gauss's law in Eq.~(\ref{eq:constraint}) is ensured by $\k\cdot\e(\k)=0$, whereas $\lambda=\pm1$ specifies the helicity corresponding to right-handed and left-handed circular polarizations.
The latter can be understood because Eq.~(\ref{eq:evolution}) for constant $v$ is solved by $\psi(t,\k)\sim e^{-\lambda iv|\k|t}$, so that the electric field evolves as $\E(t,\x)\sim\cos(v|\k|t-\k\cdot\x)\hat\e_1(\k)+\lambda\sin(v|\k|t-\k\cdot\x)\hat\e_2(\k)$ for a given wave vector.

Because of energy conservation, the complex electromagnetic field can be normalized by
\begin{align}
\int\!d\x\,v(\x)|\bm\Psi(t,\x)|^2 = 1,
\end{align}
where the integrand represents an energy density normalized by the total energy.
We now assume that the electromagnetic field forms a wave packet localized in both position and wave vector spaces.
Namely, $|\bm\Psi(t,\x)|^2$ is localized at $\x\approx\bar\x(t)$ and $|\psi(t,\k)|^2$ at $\k\approx\bar\k(t)$, where $\bar\x(t)$ and $\bar\k(t)$ are the mean position and the mean wave vector of the wave packet at a given time, respectively.
This assumption allows us to make the approximations of
\begin{align}
x^i|\bm\Psi(t,\x)|^2 &\approx \bar{x}^i(t)|\bm\Psi(t,\x)|^2, \\
k_i|\psi(t,\k)|^2 &\approx \bar{k}_i(t)|\psi(t,\k)|^2.
\end{align}
For example, the mean position under the above normalization is provided by
\begin{align}
\bar{x}^i(t) \approx \int\!d\x\,v(\x)x^i|\bm\Psi(t,\x)|^2.
\end{align}
By employing the Fourier representation in Eq.~(\ref{eq:fourier}) together with $x^i e^{\lambda i\k\cdot\x}=-\lambda i\d_{k_i}e^{\lambda i\k\cdot\x}$ and $\psi(t,\k)\equiv e^{-\lambda i\varphi(t,\k)}|\psi(t,\k)|$, we obtain
\begin{align}\label{eq:berry}
\bar{x}^i(t) \approx [\d_{k_i}\varphi(t,\k) - \A^i(\k)]_{\k\to\bar\k(t)},
\end{align}
where $\A^i(\k)=-\lambda i\e^*(\k)\cdot\d_{k_i}\e(\k)$ is the so-called Berry connection.

Amp\`ere's and Faraday's laws in Eq.~(\ref{eq:evolution}) are obtained by the principle of least action from
\begin{align}
S &= \int\!dtd\x\left[v(\x)\Psi^{i*}(t,\x)i\tensor\d_t\Psi^i(t,\x)\right. \notag\\
&\quad\left.{} - \eps^{ijk}[v(\x)\Psi^{i*}(t,\x)]\tensor\d_j[v(\x)\Psi^k(t,\x)]\right],
\end{align}
with $\Psi^*\tensor\d\Psi\equiv[\Psi^*(\d\Psi)-(\d\Psi^*)\Psi]/2$.
The equations of motion for the wave packet are then derived by regarding the action for the complex electromagnetic field as that for the wave packet~\cite{Chang:1996,Sundaram:1999}.
The Fourier representation in Eq.~(\ref{eq:fourier}) substituted into the above action leads to
\begin{align}
S \approx \lambda\int\!dt\,[\d_t\varphi(t,\k) - v(\x)|\k|]_{\x\to\bar\x(t),\k\to\bar\k(t)}.
\end{align}
The first term can be expressed as $\d_t\varphi(t,\bar\k(t))\approx\dot\varphi(t,\bar\k(t))-\dot{\bar{k}}_i(t)[\bar{x}^i(t)+\A^i(\bar\k(t))]$ in terms of the total time derivative and the mean position in Eq.~(\ref{eq:berry}).
Therefore, the action reads $S\approx\lambda\int\!dt\,L$ with the Lagrangian provided by
\begin{align}
L = \bar{k}_i(t)\dot{\bar{x}}^i(t) - v(\bar\x(t))|\bar\k(t)|
- \dot{\bar{k}}_i(t)\A^i(\bar\k(t)).
\end{align}

Finally, the Euler-Lagrange equation with respect to $\bar{x}^i(t)$ leads to
\begin{align}\label{eq:wavevector}
\dot{\bar{k}}_i(t) = -|\bar\k(t)|\d_iv(\bar\x(t)),
\end{align}
whereas that with respect to $\bar{k}_i(t)$ leads to
\begin{align}\label{eq:position}
\dot{\bar{x}}^i(t) &= v(\bar\x(t))\d_{k_i}|\bar\k(t)|
+ \varepsilon^{ijk}\dot{\bar{k}}_j(t)\B_k(\bar\k(t)) \notag\\
&= v(\bar\x(t))\frac{\bar{k}_i(t)}{|\bar\k(t)|}
- \lambda\varepsilon^{ijk}\d_jv(\bar\x(t))\frac{\bar{k}_k(t)}{|\bar\k(t)|^2},
\end{align}
where $\bm\B(\k)\equiv\d_\k\times\bm\A(\k)=\lambda\k/|\k|^3$ is the so-called Berry curvature.
The resulting two equations constitute the equations of motion for the wave packet of circularly polarized light.
It is valid as long as the variation of the potential is negligible over the spatial extension of the wave packet, requiring $|\bm\d\ln v(\bar\x(t))|\ll|\bar\k(t)|$.
The first term of Eq.~(\ref{eq:position}) together with Eq.~(\ref{eq:wavevector}) is equivalent to the null geodesic of the background spacetime as shown in Appendix~\ref{app:geodesic}.
On the other hand, the second term of Eq.~(\ref{eq:position}) transverse to both the potential gradient and the wave vector is known as the anomalous velocity, providing a correction to geometrical optics at the linear order in wavelength.
Because its sign is opposite for right-handed and left-handed circular polarizations, the trajectory of light is to acquire a dependence on its polarization, which is nothing other than the gravitational analog of the optical Magnus effect.

We note that although the consistent expression was previously derived with the Foldy-Wouthuysen transformation and the path integral formulation~\cite{Berard:2006,Gosselin:2007,Yamamoto:2017,Yamamoto:2018}, our alternative derivation allows an expanding spacetime as well as a strong gravitational potential described by Eq.~(\ref{eq:metric}).
It also clarifies that the optical Magnus effect is a purely classical phenomenon with no quantum mechanics at all.
In particular, $\hbar$ never appears unless it is introduced artificially as a bookkeeping parameter to perform the systematic expansion with respect to the wavelength.
Recently, the fully covariant equations of motion applicable to an arbitrary spacetime were derived with the Wentzel-Kramers-Brillouin approximation and were shown to be a special case of the Mathisson-Papapetrou equations~\cite{Oancea:2020,Frolov:2020,Harte:2022}.
Furthermore, the same equations of motion are expected for a wave packet of circularly polarized gravitational waves by regarding $\lambda\to\pm2$ as its helicity~\cite{Yoo:2012,Yamamoto:2018,Andersson:2021}.
A heuristic derivation with our approach applied to the linearized Einstein's equations is presented in Appendix~\ref{app:gravitational}.
We also note that a similar effect is expected for massive particles with intrinsic angular momentum but is suppressed by the mass factor~\cite{Onoda:2004}.
In the following sections, we simplify the notation as $\bar{x}^i(t)\to x^i$, $\bar{k}_i(t)\to k_i$, $|\bar\k(t)|\to k$, and $v(\bar\x(t))\to v$.

\section{Schwarzschild spacetime}\label{sec:schwarzschild}
We first apply the resulting equations of motion to gravitational lensing in the Schwarzschild spacetime with particular attention to its photon sphere and black hole shadow.
The metric in isotropic coordinates is provided by
\begin{align}
ds^2 = -\frac{\left(1-\frac{r_s}{4r}\right)^2}{\left(1+\frac{r_s}{4r}\right)^2}(cdt)^2
+ \left(1+\frac{r_s}{4r}\right)^4\delta_{ij}dx^idx^j,
\end{align}
where $r_s=2GM/c^2$ is the Schwarzschild radius and $r=\sqrt{x^2+y^2+z^2}>r_s/4$.
Then, the equations of motion read
\begin{subequations}\label{eq:schwarzschild}
\begin{align}
\frac{dk_i}{dt} &= -k\frac{x^i}{r}\d_rv, \\
\frac{dx^i}{dt} &= v\frac{k_i}{k}
- \lambda\varepsilon^{ijk}\frac{x^j}{r}\frac{k_k}{k^2}\d_rv,
\end{align}
\end{subequations}
with $v=c\,[1-r_s/(4r)]/[1+r_s/(4r)]^3$.

\subsection{Photon sphere}
The photon sphere consists of unstable circular orbits of light around a massive object~\cite{Misner:1973}.
Because
\begin{subequations}
\begin{gather}
\frac{dr}{dt} = v\frac{x^ik_i}{rk}, \qquad
\frac{dk}{dt} = -\frac{x^ik_i}{r}\d_rv, \\
\frac{d(x^ik_i)}{dt} = (v - r\d_rv)k
\end{gather}
\end{subequations}
are obtained from the above equations of motion, constant $r$ is possible only when $v=r\d_rv$.
It is achieved outside the event horizon at
\begin{align}\label{eq:photon}
r = \frac{2+\sqrt3}{4}r_s,
\end{align}
corresponding to $R=r\,[1+r_s/(4r)]^2=3r_s/2$ in Schwarzschild coordinates.
This is the radius of the photon sphere, which is found to be unaffected by the optical Magnus effect.

When $r$ is constant, $k$ is also constant because of $x^ik_i=0$.
Then, Eq.~(\ref{eq:schwarzschild}) under $v=r\d_rv$ leads to
\begin{align}
\frac{d^2x^i}{dt^2} = -\left(\frac{v}{r}\right)^2
\left(1 + \frac{\lambda^2}{r^2k^2}\right)x^i,
\end{align}
so that the orbit is provided by $x^i(t)=\cos(\omega t)x^i(0)+\sin(\omega t)\dot{x}^i(0)/\omega$ with
\begin{align}
\omega = \frac{v}{r}\sqrt{1 + \frac{\lambda^2}{r^2k^2}}.
\end{align}
This is the frequency of circular motion of light, which is found to acquire a dependence on its wavelength due to the optical Magnus effect.
In particular, the resulting frequency increases as the wavelength increases equally for right-handed and left-handed circular polarizations.

\subsection{Black hole shadow}
When a general orbit of light is considered, $E\equiv vk$ is conserved.
On the other hand, $L_i\equiv\eps^{ijk}x^jk_k$ is not conserved in the presence of the optical Magnus effect~\cite{footnote},
\begin{align}
\frac{dL_i}{dt} = \lambda\left(\frac{x^i}{r}
- \frac{k_i}{k}\frac{x^jk_j}{rk}\right)\d_rv,
\end{align}
but $L^2=r^2k^2-(x^ik_i)^2=r^2k^2\sin^2\theta$ is conserved.
Therefore, $b\equiv cL/E$ is a constant of motion, providing the impact parameter at $r\to\infty$.
The radial equation of motion can be expressed in terms of $b$ as
\begin{align}\label{eq:radial}
\left(\frac1v\frac{dr}{dt}\right)^2 = 1 - \left(\frac{bv}{cr}\right)^2,
\end{align}
where the right-hand side reaches its minimum at the radius of the photon sphere in Eq.~(\ref{eq:photon}).
When the minimum is positive (negative), the orbit of light can (cannot) cross the photon sphere, corresponding to a plunging (scattering) orbit.
The critical impact parameter separating the plunging and scattering orbits then reads
\begin{align}
b = \frac{cr}{v}\Big|_\text{Eq.(\ref{eq:photon})} = \frac{3\sqrt3}{2}r_s.
\end{align}
This is the radius of the black hole shadow~\cite{Misner:1973}, which is again found to be unaffected by the optical Magnus effect.

\begin{figure}[t]
\includegraphics[width=0.9\columnwidth,clip]{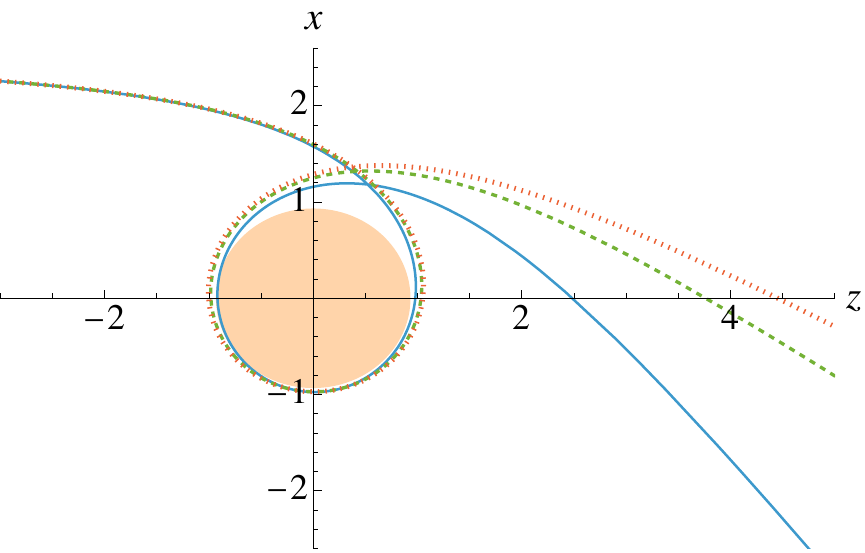}\\\medskip
\includegraphics[width=0.9\columnwidth,clip]{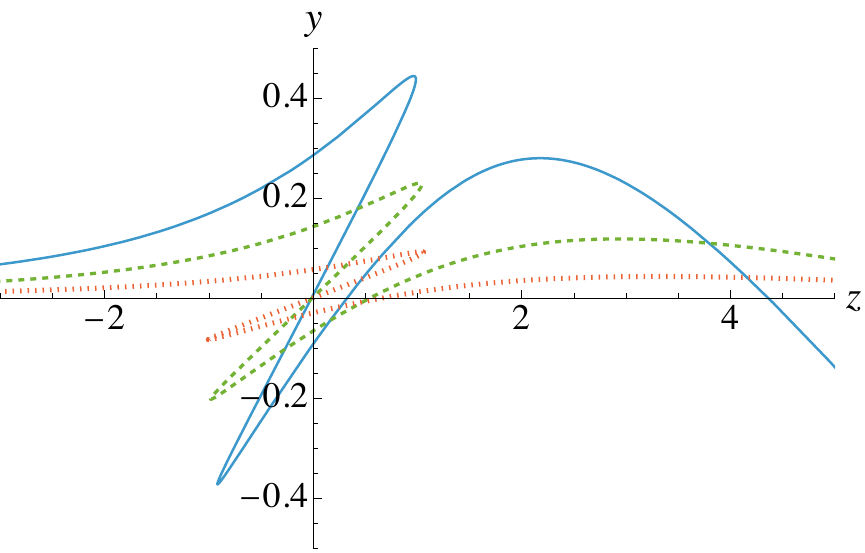}
\caption{\label{fig:trajectory}
Trajectories of circularly polarized light with $\lambda=+1$ incident from $\x\to(2.6,0,-\infty)$ in units of $r_s=1$, represented by $x$ (upper panel) and $y$ (lower panel) coordinates as functions of $z$.
The solid (blue), dashed (green), and dotted (red) curves correspond to initial wave vectors of $\k=(0,0,1)$, $(0,0,2)$, and $(0,0,5)$, respectively.
The photon sphere (orange) is also presented in the upper panel.
The corresponding trajectories for $\lambda=-1$ are obtained by the replacement of $y\to-y$.}
\end{figure}

Because the anomalous velocity does not contribute to Eq.~(\ref{eq:radial}), the radial motion is entirely unaffected by the optical Magnus effect.
However, the trajectory of light is indeed affected in three-dimensional space.
As a demonstration, three trajectories obtained by numerical integration of Eq.~(\ref{eq:schwarzschild}) are shown in Fig.~\ref{fig:trajectory} for different wavelengths but with the same impact parameter $b=2.6\,r_s$ slightly above the critical value.
Although the trajectory of light is restricted to a plane in the geometrical-optics approximation, it deviates toward transverse directions opposite for right-handed and left-handed circular polarizations due to the optical Magnus effect.
In particular, the resulting deviation increases as the wavelength increases within the validity range of $k^{-1}\ll r_s$.

\section{Weak gravitational potential}\label{sec:weak}
We then apply the equations of motion for a wave packet of circularly polarized light to gravitational lensing in an expanding spacetime with a weak gravitational potential $\phi=\phi(\x)\ll c^2$.
The metric is provided by
\begin{align}
ds^2 = a(t)^2\left[-(c^2+2\phi)dt^2
+ \left(1-\frac{2\phi}{c^2}\right)\delta_{ij}dx^idx^j\right],
\end{align}
so that the equations of motion read
\begin{subequations}\label{eq:weak}
\begin{align}
\frac{dk_i}{dt} &= -\frac{2k}{c}\d_i\phi, \\
\frac{dx^i}{dt} &= v\frac{k_i}{k}
- \frac{2\lambda}{c}\varepsilon^{ijk}\d_j\phi\frac{k_k}{k^2},
\end{align}
\end{subequations}
with $v=c+2\phi/c$.
In this section, we work up to the linear order in $\phi$ and neglect $O(\phi^2)$ without further notes.

\subsection{Deflection and transverse shift}
The above equations of motion can be solved perturbatively.
We consider a trajectory of light in the form of
\begin{align}
\x &= (b,0,ct) + O(\phi), \\
\k &= \frac{k}{c}\frac{d\x}{dt} + O(\phi) = (0,0,k) + O(\phi),
\end{align}
where $b$ is the impact parameter.
By changing the parameter from $t$ to $z=x^3$ with $dz/dt=c+O(\phi)$, Eq.~(\ref{eq:weak}) for $m=1,2$ can be expressed as
\begin{subequations}\label{eq:perturbation}
\begin{align}
\frac{dk_m}{dz} &= -\frac{2k}{c^2}\d_m\phi, \\
\frac{dx^m}{dz} &= \frac{k_m}{k}
- \frac{2\lambda}{c^2k}\varepsilon^{mn}\d_n\phi,
\end{align}
\end{subequations}
with $\eps^{mn}\equiv\eps^{mn3}$.
Then, the integration with respect to $z$ under $k_m|_{z\to-\infty}=0$ leads to
\begin{align}
k_m = -\frac{2k}{c^2}\int_{-\infty}^z\!dz'\d_m\phi,
\end{align}
and thus,
\begin{align}\label{eq:deflection}
[x^m]_{-\infty}^z = -\frac{2}{c^2}\int_{-\infty}^z\!dz'(z-z')\d_m\phi
- \frac{2\lambda}{c^2k}\int_{-\infty}^z\!dz'\varepsilon^{mn}\d_n\phi,
\end{align}
where $\phi$ in the integrands is evaluated at $\x=(b,0,z')$ and $\int_{-\infty}^zdz''\int_{-\infty}^{z''}dz'=\int_{-\infty}^zdz'\int_{z'}^zdz''$ is employed to perform the integration with respect to $z''$.
The first term provides the usual deflection of light by gravitational lensing, whereas the second term is to provide the transverse shift due to the optical Magnus effect.

When the gravitational potential is spherically symmetric, the first and second terms in Eq.~(\ref{eq:deflection}) contribute only to $x=x^1$ and $y=x^2$, respectively, because of $\d_m\phi=\delta_{m1}(b/r)\d_r\phi$.
For example, the integration with respect to $z'$ can be performed analytically for $\phi=-GM/r$, so that the trajectory of light reads
\begin{align}
x &= b - r_s\frac{\sqrt{b^2+z^2}+z}{b}, \\
y &= \frac{\lambda r_s}{kb}\frac{\sqrt{b^2+z^2}+z}{\sqrt{b^2+z^2}},
\end{align}
with $r_s=2GM/c^2$.
Therefore, the deflection angle is found to be
\begin{align}
\alpha = \left[\frac{dx}{dz}\right]_{-\infty}^{+\infty} = -\frac{2r_s}{b},
\end{align}
whereas the transverse shift is found to be
\begin{align}
\Delta y = [y]_{-\infty}^{+\infty} = \frac{2\lambda r_s}{kb}.
\end{align}
In particular, the resulting transverse shift is opposite for right-handed and left-handed circular polarizations and increases in magnitude as the wavelength increases within the validity range of $k^{-1},r_s\ll b$.

\subsection{Lens equation}
We now formulate a lens equation incorporating the transverse shift due to the optical Magnus effect.
We set the origin at the observer and direct the $z$ axis toward a source located on a plane at $z=z_s>0$.
By considering a reverse trajectory of light from the observer to the source in the form of $\x=(0,0,ct)+O(\phi)$ and $\k=(0,0,k)+O(\phi)$, the integration of Eq.~(\ref{eq:perturbation}) with respect to $z$ under $x^m|_{z=0}=0$ leads to
\begin{align}
x^m(z_s) &= z_s\frac{k_m}{k}\bigg|_{z=0}
- \frac{2}{c^2}\int_0^{z_s}\!dz\,(z_s-z)\d_m\phi \notag\\
&\quad - \frac{2\lambda}{c^2k}\int_0^{z_s}\!dz\,\varepsilon^{mn}\d_n\phi,
\end{align}
where $\phi$ in the integrands is evaluated at $\x=(x^1(z),x^2(z),z)$.
Then, by introducing two-dimensional angular coordinates on the celestial sphere as $\theta_m(z)\equiv(-1)^mx^m(z)/z$, we obtain
\begin{align}\label{eq:lens}
\theta_m(z_s) &= \theta_m(0)
- \frac{2}{c^2}\int_0^{z_s}\!dz\,\frac{z_s-z}{z_sz}\d_{\theta_m}\phi \notag\\
&\quad - \frac{2\lambda}{c^2k}\varepsilon^{mn}\d_n\phi\,\bigg|_{z=0}
+ \frac{2\lambda}{c^2k}\int_0^{z_s}\!dz\,\frac1{z_sz}\varepsilon^{mn}\d_{\theta_n}\phi.
\end{align}
The first line constitutes the usual lens equation for gravitational lensing up to the linear order in $\phi$~\cite{Schneider:1992}, which is found to be modified by the optical Magnus effect with the additional second line.

The resulting lens equation is a nonlinear integral equation for $\theta_m(z)$, which can be made tractable in the thin-lens approximation.
Namely, by assuming that the gravitational potential produced by a mass density $\rho(\x)$ is thin compared to its distances to observer and source, we regard it as localized on a lens plane at $z=z_l\in(0,z_s)$,
\begin{align}
\phi(\x) &\approx \delta(z-z_l)\int\!dz\,\phi(\x) \notag\\
&= \delta(z-z_l)2Ga_lz_l^2\int\!d\bm\theta'\Sigma(\bm\theta')\ln|\bm\theta-\bm\theta'|,
\end{align}
where $a_l$ is the scale factor at $z=z_l$ and $\Sigma(\bm\theta)=a_l\int dz\,\rho(z_l\theta_1,z_l\theta_2,z)$ is the surface mass density projected onto the lens plane.
This assumption substituted into Eq.~(\ref{eq:lens}) leads to
\begin{align}
\beta_m = \theta_m - \d_{\theta_m}\psi(\bm\theta)
+ \frac{\lambda}{k_sD_{sl}}\varepsilon^{mn}\d_{\theta_n}\psi(\bm\theta),
\end{align}
with the lensing potential provided by
\begin{align}
\psi(\bm\theta) = \frac{4G}{c^2}\frac{D_{sl}D_l}{D_s}
\int\!d\bm\theta'\Sigma(\bm\theta')\ln|\bm\theta-\bm\theta'|.
\end{align}
Here, $\bm\beta\equiv\bm\theta(z_s)$ and $\bm\theta\equiv\bm\theta(0)=\bm\theta(z_l)$ are the angular coordinates of source and image, respectively, $D_l=a_lz_l$, $D_s=a_sz_s$, and $D_{sl}=a_s(z_s-z_l)$ are the angular diameter distances from observer to lens, from observer to source, and from lens to source, respectively, and $a_s$ and $k_s=k/a_s$ are the scale factor and the physical wave number at $z=z_s$, respectively.
Therefore, gravitational lensing and the optical Magnus effect contribute to the lens equation as longitudinal and transverse derivatives of the lensing potential, respectively, whereas the latter is multiplied by the factor of $\Lambda\equiv\lambda/(k_sD_{sl})$.
It is valid in the range of $k_s^{-1}\ll\ell\ll D_l,D_{sl}$ with $\ell$ being the thickness of the lens.

\subsection{Axially symmetric thin lens}
\subsubsection{Generalities}
Possible consequences of the optical Magnus effect can be studied in more detail when the lensing potential is axially symmetric as $\Sigma(\bm\theta)=\Sigma(\theta)$.
By introducing the mean convergence,
\begin{align}
\bar\kappa(\theta) \equiv \frac{\psi'(\theta)}{\theta}
= \frac{4G}{c^2}\frac{D_{sl}D_l}{D_s}\frac1{\theta^2}
\int_0^\theta\!d\theta'2\pi\theta'\Sigma(\theta'),
\end{align}
the modified lens equation in the thin-lens approximation is expressed in a simple matrix form of
\begin{align}
\begin{pmatrix}
\beta_1 \\ \beta_2
\end{pmatrix}
=
\begin{pmatrix}
1-\bar\kappa(\theta) & \Lambda\bar\kappa(\theta) \\
-\Lambda\bar\kappa(\theta) & 1-\bar\kappa(\theta)
\end{pmatrix}
\begin{pmatrix}
\theta_1 \\ \theta_2
\end{pmatrix},
\end{align}
which relates the magnitudes of $\bm\beta$ and $\bm\theta$ by
\begin{align}\label{eq:magnitude}
\beta^2 = \left([1-\bar\kappa(\theta)]^2
+ [\Lambda\bar\kappa(\theta)]^2\right)\theta^2.
\end{align}
Each solution for $\theta$ in terms of $\beta$ predicts an image of a given source formed at
\begin{align}\label{eq:image}
\begin{pmatrix}
\theta_1 \\ \theta_2
\end{pmatrix}
= \left(\frac\theta\beta\right)^2
\begin{pmatrix}
1-\bar\kappa(\theta) & -\Lambda\bar\kappa(\theta) \\
\Lambda\bar\kappa(\theta) & 1-\bar\kappa(\theta)
\end{pmatrix}
\begin{pmatrix}
\beta_1 \\ \beta_2
\end{pmatrix}.
\end{align}
In particular, $\beta=0$ corresponds to the exact alignment of source, lens center, and observer on a single line, where the image can appear as an Einstein ring at $\theta=\theta_E>0$ satisfying $\bar\kappa(\theta_E)=1$ in the absence of the optical Magnus effect.
However, when the optical Magnus effect is incorporated, Eq.~(\ref{eq:magnitude}) no longer allows solutions at $\theta>0$ for $\beta=0$.
Therefore, the Einstein ring is found impossible to emerge, which is one of the most significant consequences of the optical Magnus effect.

The expansion $\kappa$, twist $\omega$, and shear $\gamma_m$ of an image are provided by the Jacobian matrix of the lens equation~\cite{Schneider:1992}.
From
\begin{align}
\frac{\d\beta_m}{\d\theta_n} =
\begin{pmatrix}
1 - \kappa - \gamma_1 & -\gamma_2 - \omega \\
-\gamma_2 + \omega & 1 - \kappa + \gamma_1
\end{pmatrix}_{mn},
\end{align}
we obtain
\begin{gather}
\kappa(\theta) = \bar\kappa(\theta) + \frac\theta2\bar\kappa'(\theta), \qquad
\omega(\theta) = -\Lambda\kappa(\theta), \\
\gamma_1(\bm\theta) = \left(\frac{\theta_1^2-\theta_2^2}{\theta^2}
- \Lambda\frac{2\theta_1\theta_2}{\theta^2}\right)\frac\theta2\bar\kappa'(\theta), \\
\gamma_2(\bm\theta) = \left(\frac{2\theta_1\theta_2}{\theta^2}
+ \Lambda\frac{\theta_1^2-\theta_2^2}{\theta^2}\right)\frac\theta2\bar\kappa'(\theta).
\end{gather}
The inverse of its determinant is the magnification factor and reads
\begin{align}
\mu(\theta) = \frac1{[1-\kappa(\theta)]^2 + [\Lambda\kappa(\theta)]^2
- (1+\Lambda^2)[\kappa(\theta)-\bar\kappa(\theta)]^2}.
\end{align}
Therefore, the optical Magnus effect is found to affect the twist, shear, and magnification of the image.
In particular, its twist is produced solely by the optical Magnus effect and increases in magnitude as the wavelength of light increases within the validity range.

We note that when the magnification factor is divergent at $\theta=\theta_c$, the circle with radius $\theta_c$ in the image plane and the corresponding circle with radius $\beta_c=\beta(\theta_c)$ in the source plane are referred to as a critical curve and a caustic, respectively.
Because the magnification factor can be expressed in terms of $\beta^2$ in Eq.~(\ref{eq:magnitude}) as
\begin{align}
\mu(\theta) = \left(\frac{d\beta^2}{d\theta^2}\right)^{-1},
\end{align}
$\theta_c$ and $\beta_c$ are provided by an extremum point and the corresponding extremum value of $\beta(\theta)$, respectively.
Therefore, when a source crosses the caustic, a pair of images appears or disappears on the critical curve.
In particular, we obtain $\theta_c=\theta_E+O(\Lambda^2)$ and $\beta_c=|\Lambda|\theta_E+O(\Lambda^2)$ for $|\Lambda|\ll1$, where $\theta_E$ is the Einstein radius satisfying $\bar\kappa(\theta_E)=1$.

In order to illuminate how image formation is affected by the optical Magnus effect, we consider $\theta=\theta_E+\delta\theta$ with $|\delta\theta|\sim\beta\sim|\Lambda|\ll1$, so that Eq.~(\ref{eq:magnitude}) is solved by
\begin{align}
\theta = \theta_E \pm \frac{\sqrt{\beta^2 - \Lambda^2\theta_E^2}}
{\theta_E\bar\kappa'(\theta_E)} + O(\Lambda^2).
\end{align}
Whereas two such solutions exist for $\beta>|\Lambda|\theta_E$, they merge at $\theta=\theta_E$ for $\beta=|\Lambda|\theta_E$ and then disappear for $\beta<|\Lambda|\theta_E$.
The angular coordinates where the pair of images merges are obtained from Eq.~(\ref{eq:image}),
\begin{align}
\begin{pmatrix}
\theta_1 \\ \theta_2
\end{pmatrix}_{\beta=|\Lambda|\theta_E}
= \frac{\theta_E}{|\Lambda|}
\begin{pmatrix}
0 & -\Lambda \\
\Lambda & 0
\end{pmatrix}
\begin{pmatrix}
\hat\beta_1 \\ \hat\beta_2
\end{pmatrix}
+ O(\Lambda),
\end{align}
which is a point on the Einstein radius but rotated by $\pm90^\circ$ from the direction of source in the limit of $|\Lambda|\ll1$.
In particular, the resulting rotation is opposite for right-handed and left-handed circular polarizations of light.
This is another significant consequence of the optical Magnus effect, which is to be demonstrated with simple lens models.

\subsubsection{Simple lens models}
Apart from the general discussion, the modified lens equation in the thin-lens approximation can be solved analytically for a point mass lens with $\rho=(M/a_l^3)\delta(\x)$.
The mean convergence reads
\begin{align}
\bar\kappa(\theta) = \left(\frac{\theta_E}{\theta}\right)^2, \qquad
\theta_E^2 = \frac{4GM}{c^2}\frac{D_{sl}}{D_sD_l},
\end{align}
so that Eq.~(\ref{eq:magnitude}) is solved by $\theta=\theta_\pm$ with
\begin{align}
\theta_\pm^2 = \frac{\beta^2}{2} + \theta_E^2
\pm \sqrt{\left(\frac{\beta^2}{2}+\theta_E^2\right)^2 - (1+\Lambda^2)\theta_E^4}.
\end{align}
Whereas two such solutions exist for $\beta^2>\beta_0^2\equiv2\,(\sqrt{1+\Lambda^2}-1)\,\theta_E^2$, they merge at $\theta^2=\theta_0^2\equiv\sqrt{1+\Lambda^2}\,\theta_E^2$ for $\beta=\beta_0$ and then disappear for $\beta<\beta_0$.
Because the magnification factor is provided by
\begin{align}
\mu(\theta) = \frac{\theta^4}{\theta^4 - (1+\Lambda^2)\theta_E^4},
\end{align}
the radii of the critical curve and the caustic are found to be $\theta_c=\theta_0$ and $\beta_c=\beta_0$, respectively, for the point mass lens.

Similarly, the same equation can be solved analytically for a singular isothermal sphere with $\rho=\sigma_v^2/(2\pi Ga_l^2r^2)$.
The mean convergence reads
\begin{align}
\bar\kappa(\theta) = \frac{\theta_E}{\theta}, \qquad
\theta_E = \frac{4\pi\sigma_v^2}{c^2}\frac{D_{sl}}{D_s},
\end{align}
so that Eq.~(\ref{eq:magnitude}) is solved by $\theta=\theta_\pm$ with
\begin{align}
\theta_\pm = \theta_E \pm \sqrt{\beta^2 - \Lambda^2\theta_E^2}.
\end{align}
Whereas only the upper solution exists for $\beta>\sqrt{1+\Lambda^2}\,\theta_E$, the lower solution appears at $\theta=0$ for $\beta=\sqrt{1+\Lambda^2}\,\theta_E$.
Therefore, the two solutions exist for $|\Lambda|\theta_E<\beta\leq\sqrt{1+\Lambda^2}\,\theta_E$, but they merge at $\theta=\theta_E$ for $\beta=|\Lambda|\theta_E$ and then disappear for $\beta<|\Lambda|\theta_E$.
Because the magnification factor is provided by
\begin{align}
\mu(\theta) = \frac\theta{\theta - \theta_E},
\end{align}
the radii of the critical curve and the caustic are found to be $\theta_c=\theta_E$ and $\beta_c=|\Lambda|\theta_E$, respectively, for the singular isothermal sphere.

\begin{figure}[t]
\includegraphics[width=0.9\columnwidth,clip]{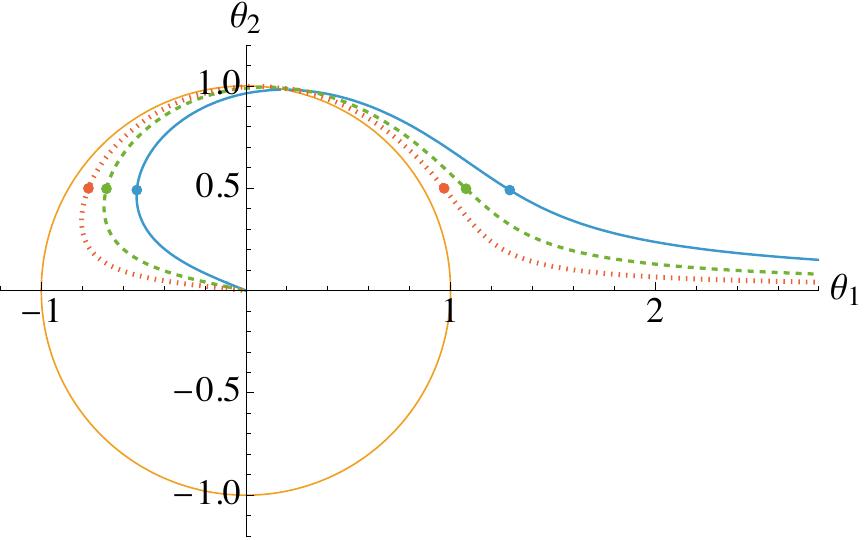}\\\medskip
\includegraphics[width=0.9\columnwidth,clip]{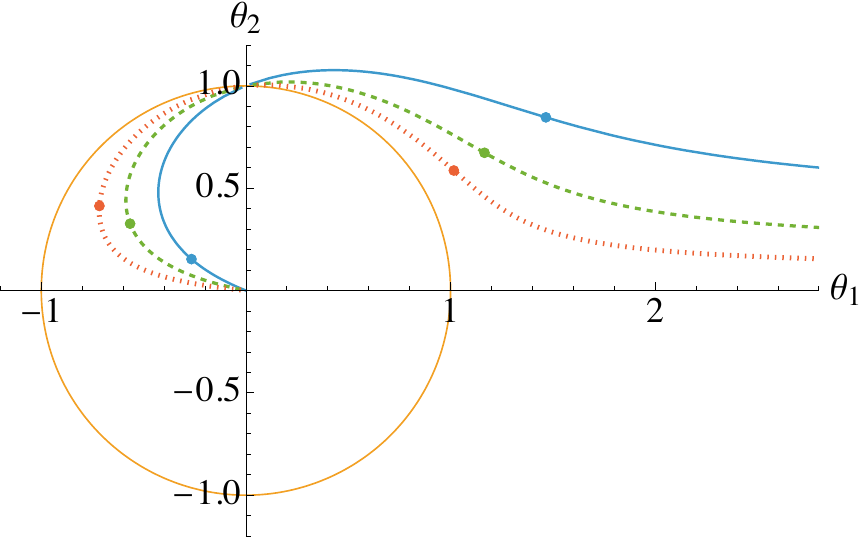}
\caption{\label{fig:image}
Angular coordinates of images formed by circularly polarized light with $\lambda=+1$ for the point mass lens (upper panel) and for the singular isothermal sphere (lower panel) in units of $\theta_c=1$.
The solid (blue), dashed (green), and dotted (red) curves correspond to $|\Lambda|=0.4$, 0.2, and 0.1, respectively, for a source located at $\bm\beta=(\beta,0)$ with $\beta$ varied over $[\beta_c,\infty)$.
A pair of images outside and inside the critical curve (orange) appears for $\beta>\beta_c$ and their locations for $\beta=2\beta_c$ are presented by dots.
They move closer to (away from) the critical curve as $\beta$ decreases (increases) and merge on the critical curve for $\beta=\beta_c$, whereas no images are formed for $\beta<\beta_c$.
The corresponding images for $\lambda=-1$ are obtained by the replacement of $\theta_2\to-\theta_2$.}
\end{figure}

The angular coordinates of each image are obtained by substituting $\theta=\theta_\pm$ into Eq.~(\ref{eq:image}), which are shown in Fig.~\ref{fig:image} separately for the point mass lens and for the singular isothermal sphere.
Although a pair of images outside and inside the critical curve appears in parallel to a source in the geometrical-optics approximation, their locations deviate toward transverse directions opposite for right-handed and left-handed circular polarizations due to the optical Magnus effect.
In particular, the resulting deviation is significant even for an infinitesimal wavelength of light when the source is located slightly outside the caustic, whereas the source inside forms no images.

\section{Summary}\label{sec:summary}
Finally, our findings are summarized as follows.
In Sec.~\ref{sec:optical}, we derived the equations of motion for a wave packet of circularly polarized light starting from Maxwell's equations in a curved spacetime.
Although the known results involving the helicity-dependent anomalous velocity are confirmed, our alternative derivation allows an expanding spacetime as well as a strong gravitational potential.

In Sec.~\ref{sec:schwarzschild}, we applied the resulting equations of motion to gravitational lensing in the Schwarzschild spacetime.
Although the radii of the photon sphere and the black hole shadow are unaffected by the optical Magnus effect, the frequency of circular motion of light is found to increase as the wavelength increases.
We also demonstrated that a general orbit of light is no longer restricted to a plane but deviates toward transverse directions opposite for right-handed and left-handed circular polarizations due to the optical Magnus effect.

In Sec.~\ref{sec:weak}, we studied gravitational lensing in an expanding spacetime with a weak gravitational potential.
By formulating the lens equation modified to incorporate the optical Magnus effect, the Einstein ring is found impossible to emerge from a point source for any axially symmetric thin lens.
This is because the optical Magnus effect makes a caustic acquire a nonzero radius so that a source inside no longer forms images near the Einstein radius.
When the source is located outside, a pair of images appears, but their locations can deviate significantly toward transverse directions even for an infinitesimal wavelength of light.

Possible consequences of the optical Magnus effect on gravitational lensing are thus illuminated as a matter of principle.
Another consequence to be reported elsewhere is circular polarization of the cosmic microwave background induced from its temperature fluctuations~\cite{Nishida:2026}.
Since they are typically suppressed by a ratio of the wavelength of light to astrophysical length scales, future work should be devoted to finding observable signatures sensitive to the optical Magnus effect.
It is also worthwhile to extend our derivation of the equations of motion to an arbitrary spacetime for a wave packet of circularly polarized light as well as gravitational waves, which may provide us with insights complementary to previous approaches~\cite{Oancea:2020,Frolov:2020,Andersson:2021,Harte:2022}.

\begin{acknowledgments}
The author thanks Teruaki Suyama for valuable discussion.
This work was supported by JSPS KAKENHI Grants No.~JP21K03384, No.~JP26K06938, and the Matsuo Foundation.
\end{acknowledgments}

\appendix
\section{Geodesic equation}\label{app:geodesic}
The geodesic equation is provided by
\begin{align}
\frac{dk^a}{ds} + {\Gamma^a}_{bc}k^bk^c = 0,
\end{align}
or equivalently,
\begin{align}
\frac{dk_a}{ds} - \frac12\d_ag_{bc}k^bk^c = 0,
\end{align}
where $s$ is the affine parameter and $k^a=dx^a/ds$ is the wave vector satisfying $g_{ab}k^ak^b=0$ for light.
In the background spacetime described by Eq.~(\ref{eq:metric}), the null condition leads to
\begin{align}
(k^0)^2 = \frac{C(\x)}{A(\x)}(k^j)^2 = \frac{(k_j)^2}{a(t)^4A(\x)C(\x)},
\end{align}
whereas the geodesic equation for $i=1,2,3$ reads
\begin{align}
\frac{dk_i}{ds} &= -\frac12a(t)^2\d_iA(\x)(k^0)^2 + \frac12a(t)^2\d_iC(\x)(k^j)^2 \notag\\
&= -\frac{(k_j)^2}{2a(t)^2A(\x)}\d_i\!\left[\frac{A(\x)}{C(\x)}\right].
\end{align}
By changing the parameter from $s$ to $t=x^0$ with $dt/ds=k^0$ and by introducing $v(\x)=\sqrt{A(\x)/C(\x)}$ and $|\k|=\sqrt{(k_j)^2}$, we obtain
\begin{align}
\frac{dk_i}{dt} = \frac{ds}{dt}\frac{dk_i}{ds} = -|\k|\d_iv(\x),
\end{align}
which is equivalent to Eq.~(\ref{eq:wavevector}).
Similarly, by employing $k^i=dx^i/ds$, we also obtain
\begin{align}
\frac{dx^i}{dt} = \frac{ds}{dt}\frac{dx^i}{ds} = v(\x)\frac{k_i}{|\k|},
\end{align}
which is equivalent to the first term of Eq.~(\ref{eq:position}).

\section{Gravitational waves}\label{app:gravitational}
Einstein's equations linearized with respect to an expanding flat spacetime
$ds^2=a^2(-dt^2+\delta_{ij}dx^idx^j)$ are provided by
\begin{align}
\d_t^2h_{ij} + \frac{2\dot{a}}{a}\d_th_{ij} - \d_k\d_kh_{ij} = 0,
\end{align}
where $h_{ij}\ll1$ is a tensor perturbation in transverse-traceless gauge satisfying $\d_ih_{ij}=h_{ii}=0$~\cite{Maggiore:2018}.
We then introduce gravitoelectric and gravitomagnetic fields as
\begin{align}
E_{ij} = -a\d_th_{ij}, \qquad
B_{ij} = a\eps^{ikl}\d_kh_{lj},
\end{align}
which are also symmetric, transverse, and traceless~\cite{Barnett:2014}.
Because of $\eps^{ikl}\d_kB_{lj}=-a\d_k\d_kh_{ij}$, we obtain
\begin{align}
\d_tE_{ij} &= -\frac{\dot{a}}{a}E_{ij} + \eps^{ikl}\d_kB_{lj}, \\
\d_tB_{ij} &= +\frac{\dot{a}}{a}B_{ij} - \eps^{ikl}\d_kE_{lj},
\end{align}
and such a pair of equations can be combined into
\begin{align}
\d_t\Psi_{ij} = -\frac{\dot{a}}{a}\Psi_{ij}^* - i\eps^{ikl}\d_k\Psi_{lj},
\end{align}
with $\Psi_{ij}\equiv(E_{ij}+iB_{ij})/\sqrt2$.
Because the first term does not allow energy conservation, we assume $\dot{a}=0$ from now on, leading to the continuity equation in the usual form of $\d_t(\Psi_{ij}^*\Psi_{ij})+\d_k(-i\eps^{kil}\Psi_{ij}^*\Psi_{lj})=0$.

The equations of motion for a wave packet of gravitational waves can be derived in parallel to Sec.~\ref{sec:motion}.
One important difference is that the Fourier representation of the complex gravito-electromagnetic field reads
\begin{align}
\Psi_{ij}(t,\x) = \int\!\frac{d\k}{(2\pi)^3}\psi(t,\k)e_{ij}(\k)e^{\lambda i\k\cdot\x}.
\end{align}
Here, $e_{ij}(\k)\equiv e_i(\k)e_j(\k)$ is defined in terms of the complex polarization vector in Eq.~(\ref{eq:fourier}), so that $\Psi_{ij}(t,\x)$ is symmetric, transverse, and traceless, and $\lambda=\pm1$ specifies the sign of helicity corresponding to right-handed and left-handed circular polarizations.
It is then straightforward to obtain the Lagrangian,
\begin{align}
L = \bar{k}_i(t)\dot{\bar{x}}^i(t) - c|\bar\k(t)| - \dot{\bar{k}}_i(t)\A^i(\bar\k(t)),
\end{align}
where $c$ is recovered and the Berry connection is provided by $\A^i(\k)=-\lambda ie_{jk}^*(\k)\d_{k_i}e_{jk}(\k)$.

We now consider a curved spacetime described by Eq.~(\ref{eq:metric}) with $a(t)=1$.
Possible corrections to the above Lagrangian include replacing $c$ by $v(\x)=\sqrt{A(\x)/C(\x)}$ as well as adding terms involving its derivatives.
Under such replacement, the Euler-Lagrange equations with respect to $\bar{x}^i(t)$ and $\bar{k}_i(t)$ lead to Eqs.~(\ref{eq:wavevector}) and (\ref{eq:position}), respectively, but with $\lambda$ replaced by $2\lambda$ because the Berry curvature is provided by $\bm\B(\k)=2\lambda\k/|\k|^3$.
Therefore, the optical Magnus effect with twice the anomalous velocity is expected for gravitational waves with circular polarizations, which should be referred to as the \textit{gravitational Magnus effect}.
In particular, all the consequences discussed in Secs.~\ref{sec:schwarzschild} and \ref{sec:weak} are applicable by regarding $\lambda\to\pm2$ as the helicity of circularly polarized gravitational waves.
However, our heuristic derivation of the equations of motion does not exclude other additional corrections at the linear order in wavelength.
How the gravitational Magnus effect is affected by the nonlinearity of strong gravitational waves is also beyond the scope of our analysis.

\end{document}